# Transformation of Traditional Marketing Communications in to Paradigms of Social Media Networking

T.R. Gopalakrishnan Nair and Kumarashvari Subramaniam

*Effective Communication for marketing is a vital field in business organizations, which is used to convey the details about their products and services to the market segments and subsequently to build long lasting customer relationships. This paper focuses on an emerging component of the integrated marketing communication, ie. social media networking, as it is increasingly becoming the trend. In 21$^{st}$ century, the marketing communication platforms show a tendency to shift towards innovative technology bound people networking which is becoming an acceptable domain of interaction. Though the traditional channels like TV, print media etc. are still active and prominent in marketing communication, the presences of the Internet and more specifically the Social Media Networking, has started influencing the way individuals and business enterprises communicate. It has become evident that more individuals and business enterprises are engaging the social media networking sites either to accelerate the sales of their products and services or to provide post-purchase feedbacks. This shift in scenario has motivated this research which took six months (June 2011 – December 2011), using empirical analysis which is carried out based on several primary and secondary evidences. The research paper also analyzes the factors that govern the social media networking sites to influence consumers and subsequently enable their purchase decisions. The secondary data presented for this research were those pertaining to the period between the year 2005 and year 2011. The study revealed promising facts like the transition to marketing through SMN gives visible advantages like bidirectional communication, interactive product presentation, and a firm influence on customer who has a rudimentary interest. It is also important to note that sampling studies mostly dealt with fortune 500 companies. It gives promising results of providing virtual shopping experience and interactive marketing through Social media networks.*

**Field of Research:** Marketing communication, Social Media Networking

## 1.0 Introduction

The marketing mixes primarily focuses on four key parameters, i.e. product, price, promotion and place. However, in this study the promotion or the Marketing Communication (MC) parameter will be discussed as it is a vital field in business organizations.

____________________________________________________________

Prince Mohammad Bin Fahd University, Aramco Endowed Chair, Prince Mohammad Bin Fahd University, Al Khobar, 31952, Kingdom of Saudi Arabia.
trgnair@yahoo.com, www.trgnair.org
Prince Mohammad Bin Fahd University, Faculty, College of Business, Prince Mohammad Bin Fahd University, Al Khobar, 31952, Kingdom of Saudi Arabia.
eiswary@hotmail.com, ksubramaniam@pmu.edu.sa

MC is used to convey the details about a business organization's products and services to the market segments and subsequently to build long lasting customer relationships. In 21$^{st}$ century, the MC platforms show a tendency to shift towards people networking which is innovative and technology bound. It is also becoming an acceptable domain of interaction. Though the Traditional Marketing Communication (TMC) channels like TV, print media, physical public relations, trade fairs etc. are still active and prominent in MC, the presences of the Internet and more specifically the Social Media Networking (SMN) has revolutionized the way individuals and business enterprises communicate. Even though at the initial stage the Internet was perceived as a "virtual announcement board" but in recent years, i.e. since 2006 or 2007 it has evolved steadily and has constructively incorporates the SMN.

The SMN sites, to name a few - such as Facebook (http://www.facebook.com), flickr (http://www.flickr.com), YouTube (http://www.youtube.com), allow individuals to be part of the same social network which allows individuals to share information and communicate on the network regardless of any geographical or timing restriction. The social media is defined as a "group of Internet based applications that build on the ideological and technological foundations of Web 2.0 and that allow the creation of exchange of User Generated Content" (Kaplan & Haenlein, 2010). Social media is a practice that takes place among identified population, who gather virtually share information, knowledge and post-purchase feedback (Safko & Brake, 2009). The emerging patterns in social networks are as friends list, status updates platform, open data access, media sharing and transparency (Forgue, 2011). The emerging numbers of SMN sites are large and consequently the numbers of users involved in it are escalating too. Figure 1 gives an overview of SMN sites with number of users for the period ending 2008. The figure indicates that millions of populations are actively engaging either in one or multiple SMN sites for the reason of "communicating".

Figure 1: Social Media Networking Sites with Number of Users for the Period Ending 2008

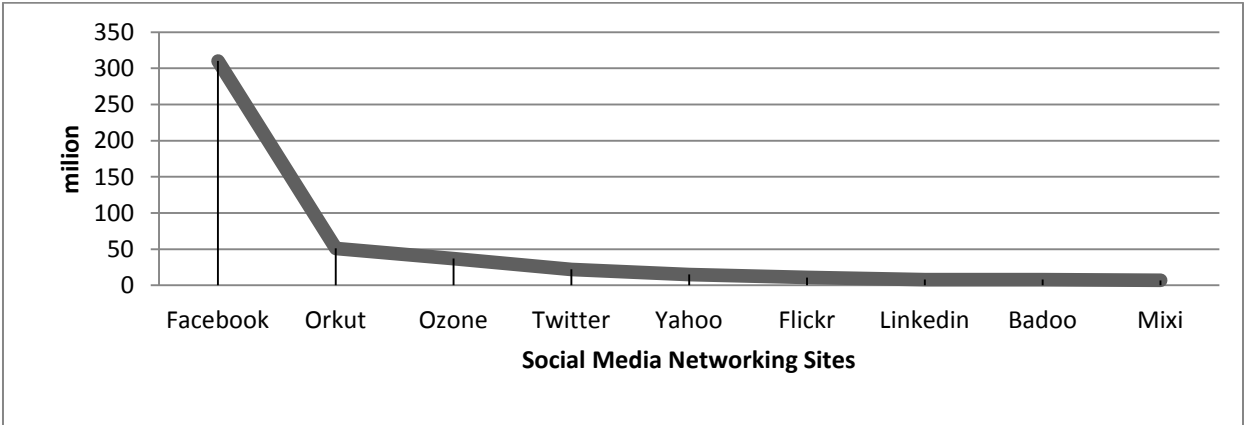

Facebook has the highest number of users with over 300 million and the number has reached over 750 million in 2011. Additionally, the other SMN sites that share the pie are SMN sites such as Orkut, Ozone, Twitter, yahoo, Flicker, LinkedIn, Badoo, Mixi, etc. The SMN sites generally have gained a strong positioning in the mind set of population due to its uniqueness such as: large market reach or penetration with low cost, easy to set up, accessibility from anywhere, quick information exchange, review, tracking and considered as "entertainment" platform. On the other hand, TMC which is a one way communication channels or platforms fails to provide that uniqueness. It is clear that the 20[th] century marketing platform remained usually unidirectional and required a substantial budget and time to get the expected outcome. For a marketing effort to be successful, or to see a substantial return on investment, it requires a high cost promotion and its ability to reach intended target audience unfortunately often remained partial. The shifting on consumer and business organization towards SMN for communication purpose is becoming clear and increasing in a large pace.

## 2.0 Factors Influencing Decisions

The rising embracement towards SMN sites has changed the way individuals and organizations interact within a business context (Barnes & Barnes, 2009).For the computer know-how generation MySpace, Mixi, Orkut etc has become the daily "checking" job. It is very much clear that SMN is affecting the way people communicate, make decisions, socialize, learn, entertain themselves, interact with each other or even do their shopping (Constantinides & Fountain, 2008). Global internet audience surpasses 1 billion visitors with Asia Pacific region accounts for 41% of internet users (Comscore, 2009). The survey conducted by (IAB Spain Research, 2009) among users of SMN sites claimed that almost 50% of the social networking services on the Internet are used as the means to obtain information. According to (Morgan, 2010) the SMN platform users have exceeded the number of email users in July 2009.

E-marketers predicted that 80% of the businesses (with more 100 employees) will adopt social media for marketing purposes by 2011 (Social Media Marketing Statistics, 2010). The favourable reception and exercise on the SMN is becomes the postmodern marketing behavior (Razorfish.com, 2008). This change in postmodern marketing is boosted up by the generation Y and the current millennium generation (Tapscott, 2009). It is clearly seen that the millennium and the generation Y are widely connected with technology and it has become as their second nature of those generations. It has becomes as an important tool in the marketing mix of promotion. The reaching ability through the SMN can reach at outsized magnitude at a small percentage of the charge and it is able to customize the desired audience groups according to the business objectives, at any time. (Caroli, 2008) 91% consumer contents is the first aid to a buying decisions. According to (Marketing Sherpa 2007) 87% trust a friend's recommendation. SMN users are 3 times more likely to trust peer opinions over advertising when making purchase decisions (Social Business Plus.com, 2010)

A number of significant factors contribute to this success. Technologies have made it possible for an individual to communicate with thousands to millions of people globally, through wireless, mobile and other connectivity (Mangold & Faulds, 2009). These virtual networks are becoming very much a part of everyday conversations in post-modern society (Simmons, 2008). As a new marketing channel, SMN sites give the easy access, relatively low organizational set-up cost, a global reach, interactive benefits etc . Business organizations are embracing SMN in order to communicate with their target audience. Additionally, SMN sites has becomes factor attribute in purchase decision making. In marketing concept such tools are important factor. According to marketing gurus the factors that influence purchase decisions are based on cultural, social, personal and psychological. However, in this current digital era the SMN sites have become another factor that influences consumers on purchase decisions. (Evans, 2008) emphasizes the role of social feedback cycle as a purchase validation tool. Social media connects these experiences back to the purchase process forming a new channel of social feedback of experiences. (Harriesinteractive.com, 2010) puts forward the fact that 71% of consumers indicate that they are substantially influenced by family and friends in arriving at purchase decision. People on Twitter recommend specific companies through their tweets for product and process and this amount to 53% of whole tweets on commercial discussions and out of that 48% express their intentions to buy certain specific products, (Mackenzie, 2011). Additionally, consumers' trusts are gained by having a secure site, reviews and web store aesthetics. According to (Channel Advisor, 2010) about 83% of holiday shoppers are influenced by consumer reviews.

Figure 2 indicates factors influencing the decision making process. In the past, consumers were influenced by type (a) factors. However, with the advancement in the technology and with the growing number of influential SMN, consumer's decision making process are not only dependent on the conventional method, but also getting empowered with SMN sites and feedbacks along with post-purchase reviews of type (b).

Figure 2: Influence of social media on purchasing decision

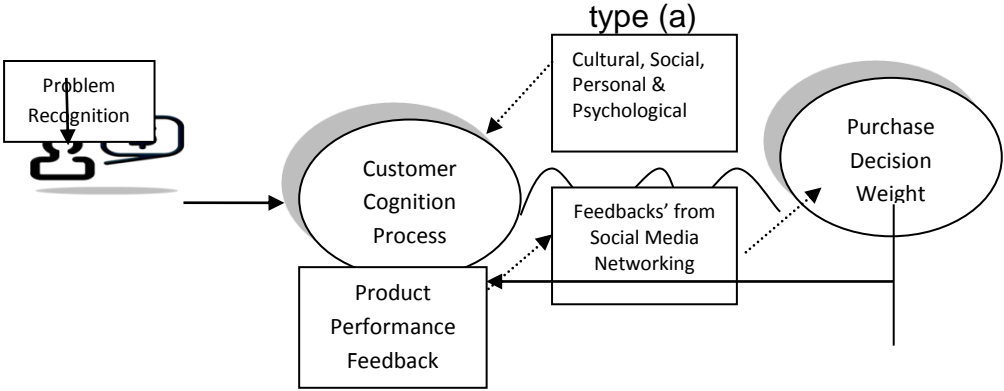

According to (Kelley & Hyde, 2002), advertising methods to target audience, steps through four typical stages: awareness, interest, desire and action (AIDA). However in the modern MC concepts it goes far beyond, short and precise communication style, heavy involvement of social communication, ability to gain product information and the controlling mechanism. (Erickson, 2010) indicates that the millennium generation playing the leading role in social network usage.

Figure 3: (Kathryn, 2010), Growth of Social Networking Site's Usage, for the period of 2008-2010, by Generation

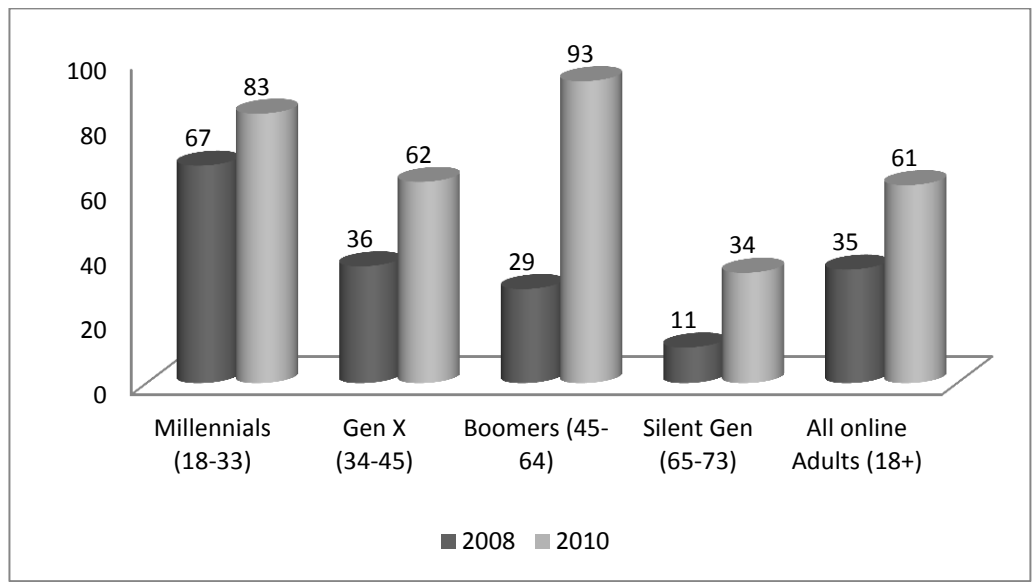

The Figure 3 indicates the millennium generation has the highest social networking rate. The rate among Generation X increased about double the percentage, while it tripled among the Silent Generation, going from 11% to 34%. Overall, 79% of US adults above 18 years can be classified as "online." The diagram also clearly indicates that a solid majority of working adults are online regularly. However, adults in their retirement years are online at notably lower levels. No one can deny that videos, messages, online advertisements are reviewed every milliseconds and it reflects that this fast growing SMN sites are well connected and getting integrated with people's life, (Lichtenberg, 2011). The thought for SMN as a communication tool was experimented in the last couple of years. But what really happening now is, the SMN is meeting or exceeding the customer's perceptions in providing hybridized information what so ever they needed virtually. This progression will challenge business organization how they want to prepare virtually or involve on the SMN in order to be connected with people. Moreover, the SMN has become a part of life and its growth is someway subtle. It is to be noted that  SMN is growing steadily and continuously.

## 3.0 Ultimate Virtual Shop with decisions at home

The Ultimate Virtual Shop (UVS) intends to give its target audience the pleasure of having the mall viewed on the personal computer, iPods, or tablet.  The UVS is going to be a success in this digital era,  as it provides the digital generation unique benefits like transparent communication, learning opportunity before action and post-purchase feedback from other consumers in the virtual arena etc. which are  absent on a regular mall shopping experience. Therefore, the reality of the virtual shop not only gives the benefit of saving time on travelling, hitting the traffic, long queues and so on. Ultimately, it gives the value of freedom, customize and personalize touch. Finally, everything happens fast. Tesco Corporation has successfully captured its South Korean market in the UVS. The country that has 50 million population and about a fifth of them possessing smart phones and has long working hours and less leisure time habit has naturally turned to these new modes of markets.

Survey by (Harrisinteractive.com, 2010) claimed that 90% of consumers have a better overall shopping experience when they research products online before shopping in-store. The number of business organization engaging in the SMN sites is rather increasing. It is a clear indication that social media can be viewed as an important channel and tool to interact with target audience and a tool for creating brand awareness too. When consumers stay at home and engage in purchasing process, this actually leads to a budget bound operation, as the consumers are not walking through the mall leading to an impulse purchase. Additionally, the demographic pattern of woman consumers is changing in a dramatic pace because today, women are working longer hours and the nature of traditional family pattern is changing. Time is precious and people are looking forward to more convenience. Hence people tend to decide their further purchases of food and other stuff virtually while waiting for a bus, tube etc.

## 4.0   Outreach Analysis

SMN has become increasingly noticeable as a virtual advise seeking platform. It is given the freedom for people in decision making. The SMN sites have attracted a large number of individuals that contribute various contents on the internet, (Bell, 2007). On the hand, business enterprises have made huge investments in recent years in order to gain high return on marketing investments. To start with Microsoft - 240 millions in Face book platform and Google - 900 millions in MySpace. Additionally, other largest Fortune 500 firms use Twitter, Facebook, YouTube or blogs to heavily dialogue with their target audience, 79% business organizations are accounted for these transactions (Thajudeen, 2011).

Figure 4: (Beese, 2011), Business Industries Accounted for Most Facebook Pages

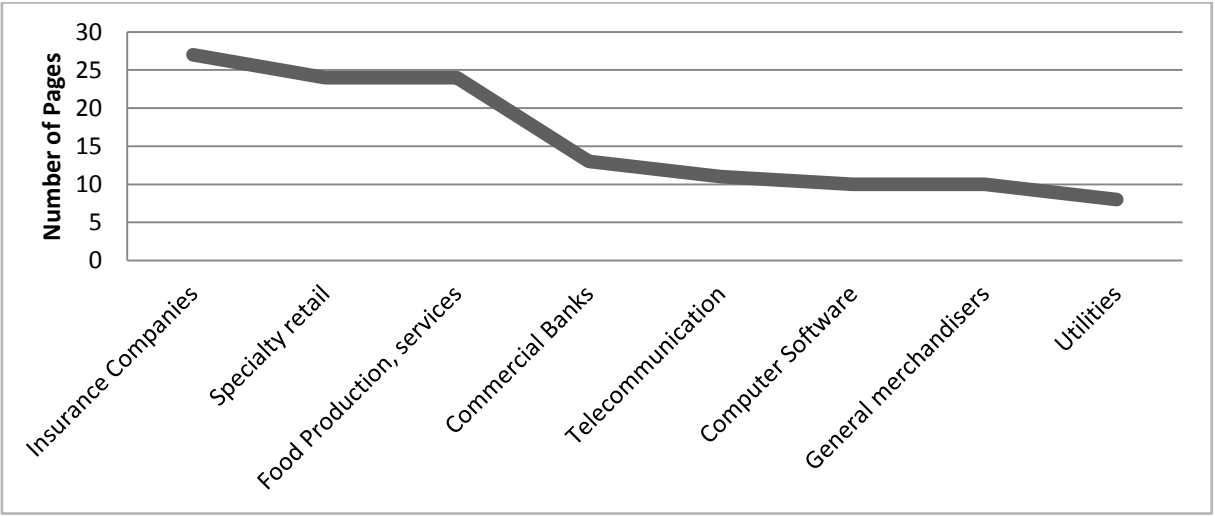

Figure 4 shows that insurance companies accounted with most facebook pages, with 27 followed by specialty retail and food production & services which both accounts for 24 pages. On the other hand, the most effective platforms in terms of mobilizing consumers to talk about products are Facebook with 86% followed by Twitter at 65%, blogs and reviews are tied at 55%.

Figure 5: Companies Using Social Media Networking Sites

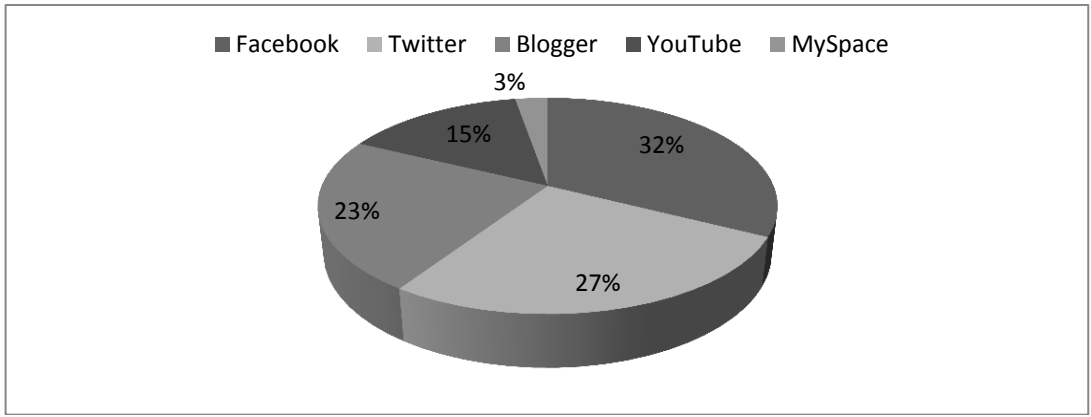

The pie-chart, Figure 5 shows that most business enterprises use Facebook and followed by twitter, blogs, Youtube and MySpace in order to communicate with their target market. In addition to that, 308 out of the 500 Fortune have an active Twitter account, and nearly half of the whole tweets are done by top 200 companies (Dugan, 2011). The companies that ranked higher in the Fortune 500 appear to be embracing Twitter more intensely than those that are ranked lower. Further, this study shows that Google, ranked as number 92, has maximum amount of followers of all the Fortune 500, with roughly over 3 million followers. Whole Foods, Starbucks and Southwest Airlines etc have followers amounting to a million. The Figure 6 indicates Fortune 500 business enterprises that have the most FaceBook fans, (Cohen, 2011)

Figure 6: Companies with Number of Facebook Fans

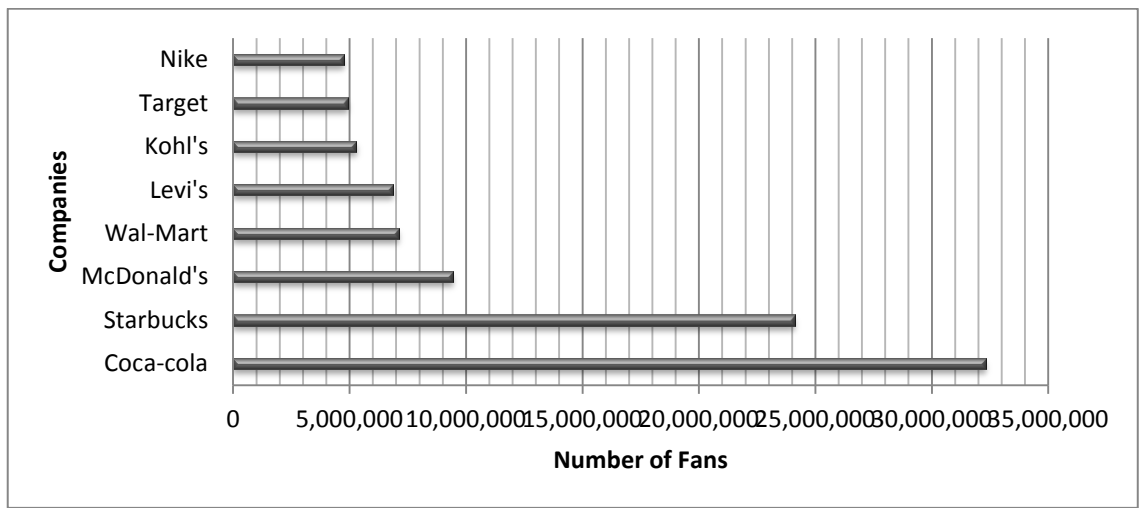

SMN is not only used as a social networking platform but its usage is moving towards doing business virtually (Benioff, 2011). SMN plays a key role in the marketing mix of promotion, where it's have a rich ability to interact with people and effectively connecting to its market. SMN sites are the fastest-mounting class on the Internet and doubling-up their traffic over the last year (Comscore.com, 2009). In addition to that, in the US social media and blogs reach nearly 80% of active US internet users and represent the majority of American's time online. Out of those 80% population 60% of people use to learn a specific brand or information about specific retailer. It is also noticed that 48% of these consumers responded to a retailer's offer as posted on FaceBook or Twitter.

Figure 7: (Henrikson, 2011), US Internet Users who use Social Networking Sites – by Age

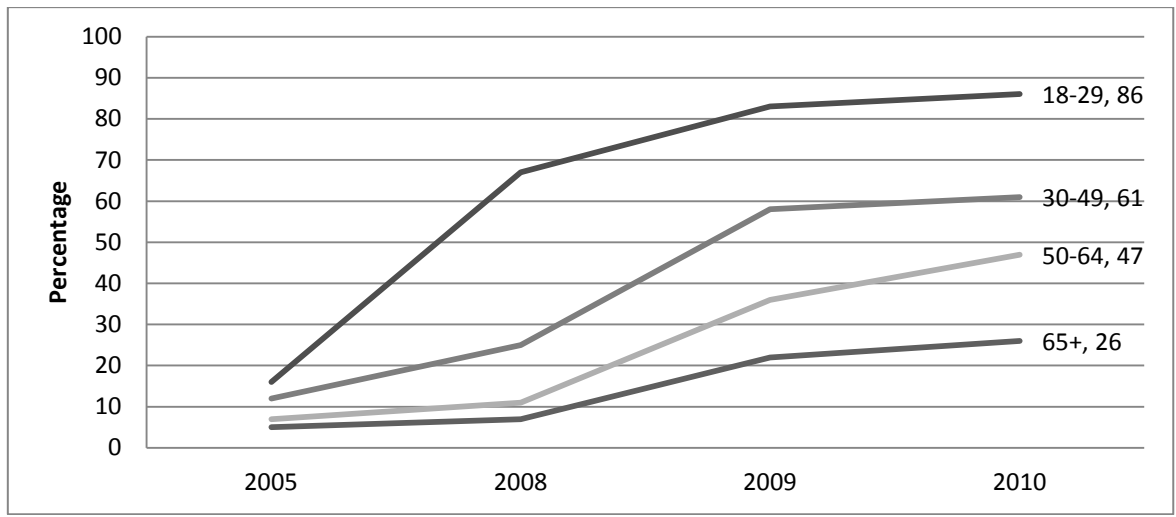

(Forrester Research, 2010), has stated that sales through the Internet will be accounted for 53% of total retail sales, as consumers actively and increasingly use the Internet to research products before a purchase decision is made. Further, Forresters also projected that the total retail sales for each year, beginning from 2010 to 2014 will be influenced by the World Wide Web at a rate of 46%, 48%, 50%, 51% and 53% for each year respectively.. (Bazaarvoice.com, 2010) found that 91% of millionaires always look at reviews before they buy luxury goods and 68% of ultra-affluent shoppers tend to use consumer reviews for finalizing purchases.

## 5.0 Conclusion

The Social Media Networking (SMN) sites are becoming more effective in marketing information dissemination and it is found that the percentage of such paradigm shift will increase continuously in the near future. SMN sites or platforms are not a direct business solution approach but it is another tool that can be used to build and maintain relationships and increase returns on the marketing investments business firms do. The progression of SMN sites and the involvement of consumers with them, are growing in fast pace. Therefore, it looks logical for modern business organizations to come forward to a point to involve themselves appropriately in the SMN sites, in order to engage with their targeted evolving market segments. They have to further study how they can increase their returns on marketing investment using these trends and enable themselves to generate a long lasting customer relationship. This scenario might compel business organizations to arrive at a trade-off position between TMC platforms and SMN, in order to satisfy their target audience and sustain their business surplus. In conclusion, there exists a good chance that most of the business organizations may embrace the SMN which will give them a visible advantage in bidirectional communication, interactive product presentation, and a firm influence on customer who has only a rudimentary interest in the beginning. Finally, it can expose promising features of providing virtual shopping experience and interactive marketing.